\documentclass[oribibl]{llncs}

\usepackage{amsmath}
\usepackage[colorlinks=true, linkcolor=blue, citecolor=blue, urlcolor=blue]{hyperref}
\usepackage{balance}
\usepackage{graphicx}
\usepackage{wrapfig}
\usepackage{booktabs}
\usepackage{tabularx}
\usepackage{multirow}
\usepackage{url}

\usepackage{fancyhdr} 
\fancypagestyle{firststyle}
{
\fancyhf{}
\fancyfoot[C]{\scriptsize{Proceedings of the 15th EAI International Conference on Digital Forensics \& Cyber Crime (EAI~ICDF2C 2024), Dubrovnik, Springer, 2025, pp. 186--211. This version is the authors' copy. The publisher's definite version is available via \url{https://doi.org/10.1007/978-3-031-89363-6_11}.}}
}

\begin{document}

\title{What Do We Know About the \\ Psychology of Insider Threats?}
\author{Jukka Ruohonen\orcidID{0000-0001-5147-3084} \and Mubashrah Saddiqa \\
  \email{\{juk, msad\}@mmmi.sdu.dk}}
\institute{University of Southern Denmark, S\o{}nderborg, Denmark}

\maketitle

\begin{abstract}
Insider threats refer to threats originating from people inside
organizations. Although such threats are a classical research topic, the
systematization of existing knowledge is still limited particularly with respect
to non-technical research approaches. To this end, this paper presents a
systematic literature review on the psychology of insider threats. According to
the review results, the literature has operated with multiple distinct theories
but there is still a lack of robust theorization with respect to psychology. The
literature has also considered characteristics of a person, his or her personal
situation, and other more or less objective facts about the person. These are
seen to correlate with psychological concepts such as personality traits and
psychological states of a person. In addition, the review discusses gaps and
limitations in the existing research, thus opening the door for further
psychology research.
\end{abstract}

\begin{keywords}
cyber crime, organizational security, security incident, inside job, deterrence,
personality traits, dark traits, systematic literature review
\end{keywords}

\section{Introduction}

\thispagestyle{firststyle} 

Insider threats refer to perceived or real threats that come from people inside
organizations. These people, the insiders, such as employees and contractors,
have inside information about an organization's security policies and practices,
including cyber security measures. Due to their work, they also have legitimate
access to organizations' information systems, computer networks, and other
information technology infrastructures. Therefore, insider threats are
particularly difficult to detect and prevent. If an insider threat realizes, the
consequences may include theft of confidential or even classified information,
theft of intellectual property and trade secrets, sabotage of information
systems, or more general fraud. Despite such serious consequences, insider
threats have been common throughout the world. According to some industry
reports, even more than a half of all cyber attacks have been conducted by
insiders~\cite{Mathews17}. As soon discussed, such numbers may be partially
explained by the collusion between insiders and external threats, such as is the
case with phishing. Thus, other studies indicate smaller numbers. For instance,
according to media sources, about ten percent of all data breaches have involved
insiders~\cite{Ruohonen24IWCC}. Whatever the actual numbers may be, it can be
concluded that insiders pose a significant threat to most organizations. That
said, the issue is nontrivial and problematic because insiders are also a
valuable asset to organizations. No organization can function without~people.

Before continuing any further, it should be understood that the research on
insider threats has been strongly divided between technical and non-technical
research approaches~\cite{Elmrabit20}. The technical branch of research has
focused on different profiling and anomaly detection techniques for computer use
and network traffic. Thus, this branch aligns with the nowadays popular zero
trust security model, which assumes that a part of an organization's
technological infrastructure has already been compromised; therefore, logging,
profiling, fine-grained access controls, and other related techniques should be
applied~\cite{CISA23}. Some studies have also connected technical profiling to
insiders' cognitive styles~\cite{Santos12}, but such bridge building studies
have been rare. In contrast, the non-technical breach of research has focused on
human behavior within organizations, often drawing from sociology, criminology,
and related social science fields. Also psychology has been a common reference
field. The divide in the research has also caused some schisms. While some
authors have argued that the non-technical branch has often been
downplayed~\cite{Sarkar10, Sokolowski16}, others have argued for a balanced
approach that takes both technical and non-technical factors into
account~\cite{Zaytsev17}. For the present purposes, it is important to emphasize
that the non-technical research branch is generally less known than the
technical branch of research.

Therefore, this paper presents a systematic literature review on the psychology
of insider threats. While there are existing systematic literature
reviews~\cite{Gheyas16}, as well as more general review
articles~\cite{Marbut24}, the systematization of existing knowledge is still
limited particularly with respect to psychology. The paper's contribution is
thus clear and welcome. In what follows, the paper proceeds in a straightforward
manner: the methodology for the systematic literature review is addressed in
Section~\ref{sec: methodology}, the results of the review are presented in
Section~\ref{sec: results}, and the conclusion follows in the final
Section~\ref{sec: conclusion}. As discussed in the opening section, out of
necessity, the review does not rely on quantification or related systematization
techniques but instead concentrates on pinpointing relevant insights and gaps in
the existing knowledge. The gaps and associated problems provide good
opportunities for further research, as elaborated in the concluding~section.

\section{Methodology}\label{sec: methodology}

Conventional guidelines were followed for the systematic literature review; a
protocol was specified prior to the literature search and only well-known
databases were used to retrieve peer reviewed scientific
papers~\cite{Nightingale09}. Eight academic databases were queried, including
the major ones in computer science; ACM Digital Library, IEEE Xplore, Springer
Link, and ScienceDirect. Otherwise, the search was kept plain and simple;
auxiliary techniques~\cite{Kitchenham13} were thus omitted. For instance,
bibliometric details (such as citations or impact factors) were not considered,
and the search was not extended toward papers cited in the papers obtained from
the database queries. The following query string was used to retrieve the search
results from each database on June 10, 2024:
\begin{align*}
\notag
&(\textmd{insider}~\textmd{AND}~\textmd{threat}~\textmd{AND}~\textmd{psychology})~\textmd{OR}~\\ \notag
&(\textmd{insider}~\textmd{AND}~\textmd{threat}~\textmd{AND}~\textmd{psychological})
\end{align*}

\begin{figure}[th!b]
\centering
\includegraphics[width=11cm, height=8cm]{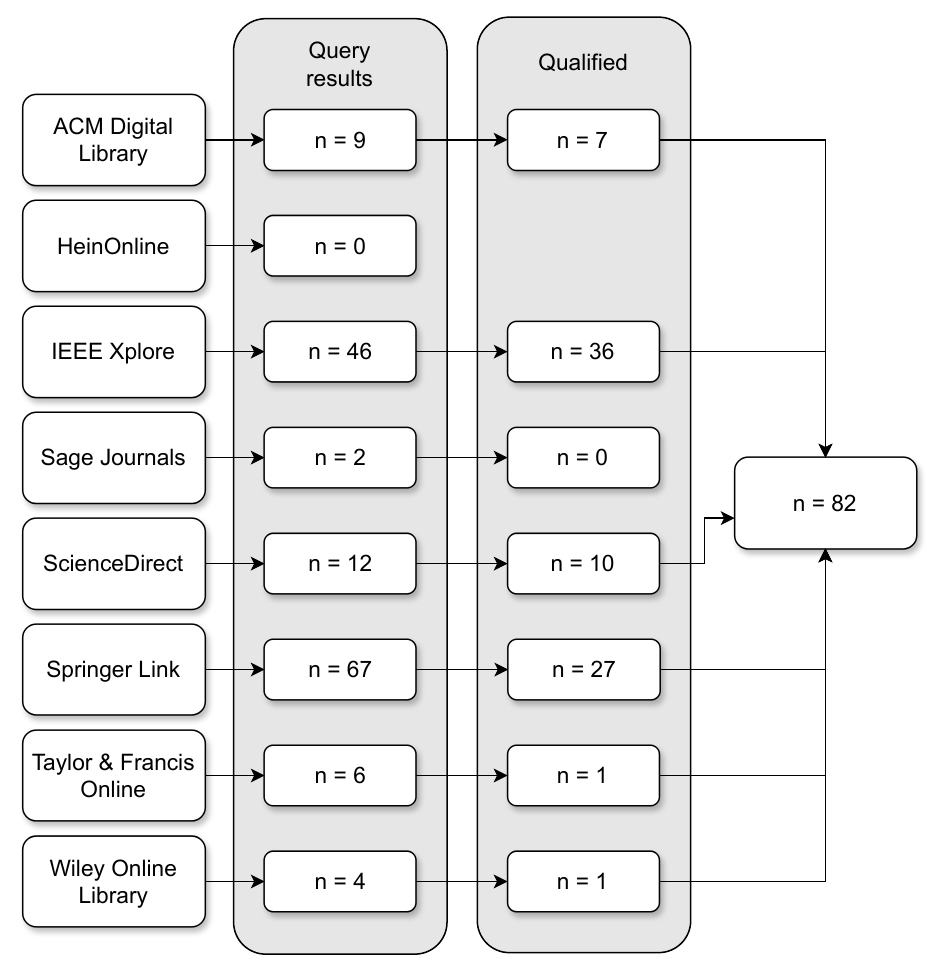}
\caption{The Literature Search}
\label{fig: sample}
\end{figure}

The search was limited to abstracts for all databases except Springer Link,
which does not allow specifying the search location. For this database, the
search term ``insider threat'', as specified in quotation marks, was used
together with the terms psychology and psychological, as specified above. In
addition, the searches were restricted to conventional articles in journals and
conference proceedings, such that book chapters and related content were
excluded. As can be seen from Fig.~\ref{fig: sample}, in total nearly 150 papers
were returned by the searches. This large amount was reduced by qualifying only
scientific papers, such that editorials and related content were excluded, which
dealt with insider threats and psychology in the main body of the papers'
text. This exclusion criterion implied that those papers were excluded that only
dealt with the subject matter in terms of literature references, for
instance. The final sample for the literature review contains $n = 82$ peer
reviewed papers. Most of these are computer science papers. Unlike what was
expected prior to the searches, only a few relevant social science papers were
retrieved. Moreover, HeinOnline, a major search portal for legal research, did
not return a single paper.

The search procedure satisfies two of the desired properties of systematic
literature reviews; the procedure was structured and it is transparent. However,
the third desired criterion, comprehensiveness, is not perfectly fulfilled
because particularly the restriction to abstracts may have excluded some
relevant papers. Even with this restriction, the amount of papers is still much
larger than in previous systematic literature reviews ($n =
37$)~\cite{Gheyas16}. Furthermore: while some authors have maintained that
\textit{all} research addressing a specific question should be
included~\cite{Nightingale09}, others have been less strict, arguing that all
\textit{relevant} research should be covered~\cite{Hiebl23}. In terms of
relevance, the sample can be argued to be sufficient for addressing the existing
knowledge about the psychology of insider threats---even if some papers are
missing. As will be seen, the existing knowledge is still fairly immature, and
this overall picture of immaturity would not likely change with a few more
papers added to the review. Therefore, it could be also said that the paper is a
systematic mapping study instead of a systematic literature
review~\textit{per~se}. In general, systematic mapping studies concentrate on
thematic analysis, identifying relevant insights and research gaps, while the
latter focuses on systematization of existing evidence~\cite{Petersen08}. Due to
various different theories, variables, methods, and datasets, the insider threat
literature is unfortunately too diverse for deducing about actual evidence in
the form of a meta-analysis or some other quantitative review technique.

\section{Results}\label{sec: results}

\subsection{Taxonomies}

Many of the papers presented or dealt with different taxonomies for insider
threats. Because these taxonomies are to some extent related to psychology, it
is helpful to consider some examples from the literature. The examples also
help at a high-level analytical framing of the literature.

Thus, the taxonomies typically separate intentional and unintentional insider
threats and incidents~\text{\cite{Greitzer14, Lachen20, Prabhu22}}. The actual
labels used tend to vary slightly from a study to another but the same theme is
still present; the same separation can be thus also referred to with the terms
malicious and non-malicious~\cite{Carroll14, Harms22, Marbut24}. What separates
the two is \textit{intention}; a malicious insider intends to compromise an
organization for some \textit{goals}, whereas non-malicious insiders may
unintentionally compromise the organization with accidental or negligent but
well-meaning mistakes or errors. With this basic separation at hand, it is
possible to continue to further insider types. For instance, accidental insiders
have no intent to cause harm, negligent and mischievous insiders cause harm but
have no malicious intent, and purely malicious insiders cause deliberate harm
for plain malice or other goals~\cite{Prabhu22}. A further option is to consider
incidental and deliberate harms. Then, whistleblowers may be seen to cause
incidental harm for some societal goals and misbehavers for some personal
reasons, whereas malicious insiders cause deliberate harm together with
ideologues who have some political ideals and goals
thereto~\cite{Renaud24}. This taxonomy is illustrated in Fig.~\ref{fig:
  taxonomy}, which further operates with an awareness of organizations' security
controls and compliance~procedures.

\begin{figure}[th!b]
\centering
\includegraphics[width=\linewidth, height=8cm]{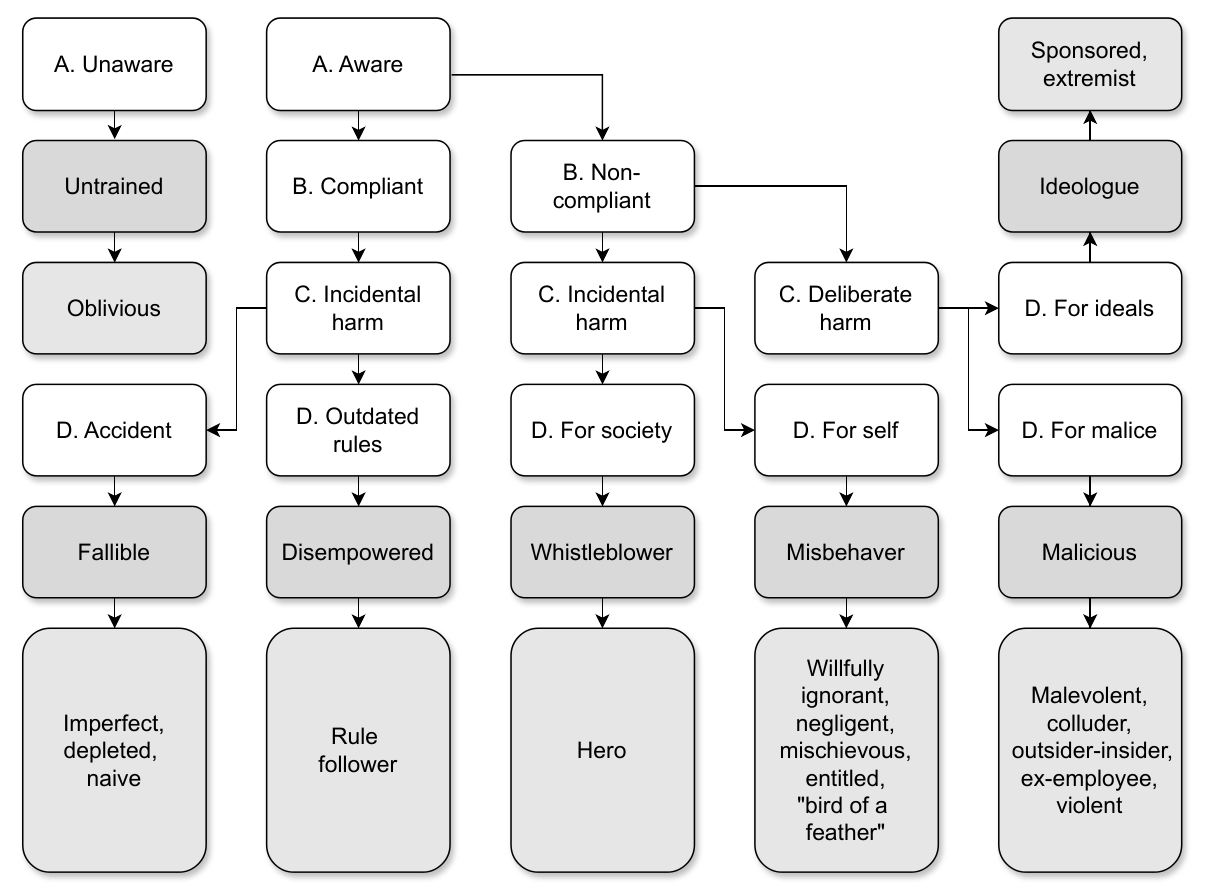}
\caption{An Example Taxonomy from the Literature (adopted from \cite{Renaud24})}
\label{fig: taxonomy}
\end{figure}

With respect to computer science research in particular, relevant is also the
collusion between internal and external threats. According to the literature, a
typical example would be social engineering with which external threat actors
lure insiders to compromise an organization or parts of it. The example is
relevant in terms of psychology research because it pinpoints toward analyzing
the psychological vulnerabilities of insiders together with the psychological
persuasion and manipulation techniques of external threat
actors~\cite{Uebelacker14}. In terms of taxonomies, on the other hand, the
collusion leads to further typologies, such as pure insiders, insider associates
(such as contractors, security guards, or cleaners), inside affiliates (such as
spouses, friends, or clients), and outside affiliates, including former
employees~\cite{Sarkar10}. The literature has also considered other types of a
collusion. For instance, a concept of ``cyber friendly fire'' has been used to
describe situations in which intentional security operations intended to protect
an organization cause unintentional harms to the organization's
security~\cite{Carroll14}. Another question is the usefulness of the taxonomies
to begin with.

The insider threat taxonomies may be useful for organizations in terms of risk
analysis. When knowing an organization's valuable assets, it possible to better
analyze typical and plausible insider threats. As assets vary from an
organization to another, so do the insider threats. For instance, it has been
suspected that some intentional data leaks from central governments have
involved insiders with political goals~\cite{Ruohonen24PRR}. Such insiders would
be classified into ideologues. The assets of most multinational companies are
entirely different than politically relevant documents, and thus also the
insider threats are typically quite different. In terms of empirical research,
however, it remains unclear how useful the taxonomies are in practice. The
insider types might be useful in criminology research, for instance, but the
problem is that there are no easily and publicly available datasets for caught
insiders from organizations themselves, law enforcement, or other public
authorities. Yet, in terms of theoretical research, the taxonomies are
analytically useful already because intentions and goals connote with
\textit{motives}. Together, intentions, goals, and motives (or desires)
constitute a formidable conundrum in philosophy and moral
psychology~\cite{Chan08}. However, it is unnecessary to delve deeper into this
problem area already because the insider threat literature sampled is not
theoretically rich in this regard. It is typically assumed that malicious
insiders are more or less rational in their actions.

\subsection{Theoretical Foundations}\label{sec: theory}

Motives of malicious insiders constitute a central theoretical tenet in the
literature. In particular, the so-called \textit{situational crime prevention}
(SCP) theory posits that a crime occurs because of two factors: a motive and an
opportunity~\cite{Azaria14}. Then, a crime can be often prevented simply by
removing either factor from the equation. In terms of insider threats and cyber
security, a motive might be countered by rigorous logging, monitoring, and
auditing, which help to hold culprits accountable, while opportunities might be
reduced by fine-grained authentication and authorization procedures, strict
access controls, and other related defensive cyber security
measures~\cite{Koien19, Safa19}. Together, such technical solutions should
increase the risk of getting caught, the effort required for an insider to
commit an offense, and the probability of obtaining a reward from the
offense. In terms of reducing the probability of obtaining rewards, many
additional techniques can be implemented; among these are digital signatures and
watermarking, information and hardware segregation, encryption, automatic data
destruction schemes, and minimization of reconnaissance
information~\cite{Safa18}. While these techniques may not fully prevent an
insider incident from happening, they should still discourage employees or other
associated people from misconduct.

These preventive techniques make the SCP a close associate of another
criminological theory, the so-called \textit{general deterrence
  theory}. According to this theory, criminals make decisions based on the
perceived utility (or benefits) of their actions and the costs (or sanctions)
involved. Increasing the costs is achieved by deterrence, meaning that an
organization should at least invest in education and outreach to inform
employees or other relevant people about the penalties from
misconduct~\cite{Azaria14}. Deterrence is essentially about external factors,
like many other organizational measures, such as monetary rewards from a job
well done, but psychologically also intrinsic factors are
relevant~\cite{LeeLallie23}. Traditionally, intrinsic motivation refers to an
inherent desire to undertake work even without specific rewards, but the concept
can be also extended toward the insider threat context. For instance, an
employee's intrinsic motivation to comply with an organization's security
policies may be weakened by poor job satisfaction, among other things.

\begin{wrapfigure}{r}{0.45\textwidth}
\centering
\includegraphics[width=\linewidth, height=3cm]{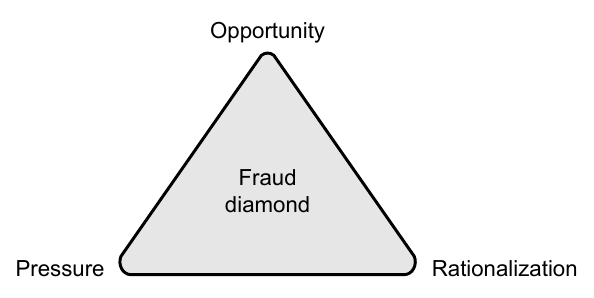}
\caption{The Fraud Diamond}
\label{fig: fraud}
\end{wrapfigure}

Furthermore, to some extent, the SCP aligns with a \textit{rational choice}
viewpoint to crime according to which crimes are deliberate; a motive to commit
a crime correlates with an intention of obtaining some reward, whether financial
resources or material goods, prestige, excitement, or something
else~\cite{Sokolowski15, Willison06}. This viewpoint is implicitly present also
in the popular so-called fraud diamond or triangle (see Fig.~\ref{fig: fraud})
that contains three angles corresponding with a motive (or a pressure), an
opportunity, and rationalization~\cite{Farahmand13, Harrison18, Mekonnen15,
  Othman22}. The rationalization refers to the rational means by which criminals
justify their actions to themselves. Oftentimes, such a justification involves
contemplating about a reward. For instance, an indebted insider might rationally
justify his or her offense due to a fact that he or she needs to pay his or her
bills. Rationalization may apply also to non-malicious insiders. For instance, a
person might share a password with a colleague because he or she rationalizes
that no one cares about such a supposedly minor violation of an organization's
security policy. To counter such rationalization and associated excuses,
organizations should seek to have clear documents on policies and their
enforcement~\cite{Farahmand13, Kisenasamy22, Safa19, Safa18}. These include also
guidelines on ethical conduct, intellectual property, and trade secrets, among
other things. There is also existing research on such policies implemented in
large information technology companies~\cite{Dong24}. Despite the name of the
rational choice viewpoint, it is usually assumed that criminals still operate
only in terms of so-called bounded or limited rationality; they do not have
perfect information available in the environment in which criminal
decision-making takes place~\cite{Willison06}. To this end, some authors have
considered thresholds for decision-making involving an intention to commit a
crime based on judgments from limited information cues~\cite{MartinezMoyano06,
  Sokolowski15, Sokolowski16, Sticha16}. In addition to a lack of information,
psychological and related factors presumably further reduce rationality and
rational decision-making.

\begin{figure}[th!b]
\centering
\includegraphics[width=11cm, height=5cm]{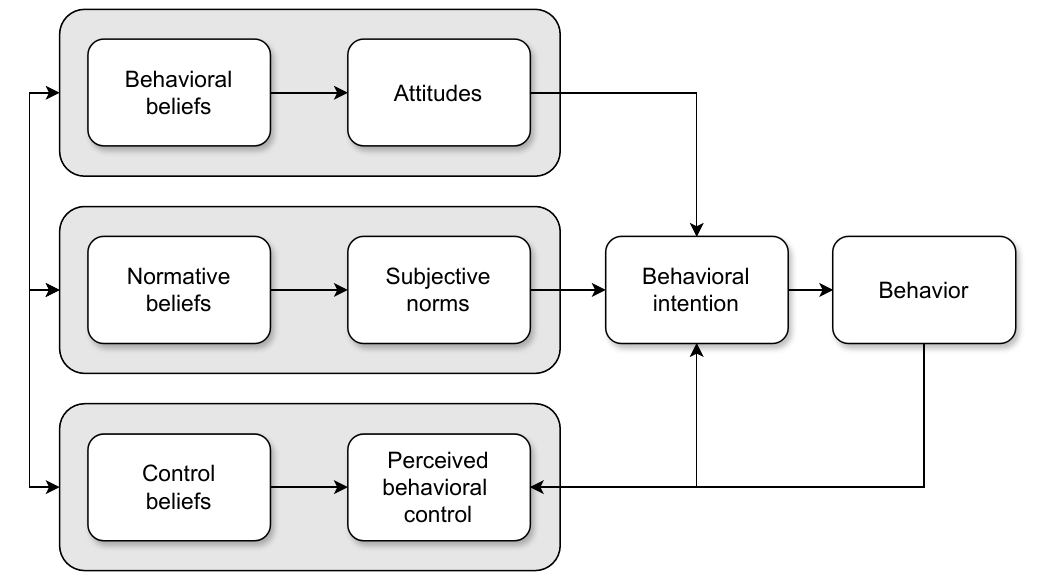}
\caption{The TPB in Essence (adopted from \cite{Lachen20})}
\label{fig: tbp}
\end{figure}

The \textit{theory of reasoned action} (TRA) is a popular general theory for
explaining human behavior~\cite{Fishbein77}. It posits that behavioral
intentions, which motive actual behavior, are composed of two factors: attitudes
and subjective norms. The former are shaped by the expected outcomes of a given
behavior, while subjective norms include the perceived social pressure to
perform or not to perform a given action. These tenets correlate with the
earlier points about the other theories; expected outcomes may refer to the
rewards from an offense, for instance, while social pressure and other
subjective norms may include knowledge about deterrence measures. The insider
threat literature has considered also an extension to the TRA, the so-called
\textit{theory of planned behavior} (TPB). It augments the TRA with perceived
behavioral control, meaning the difficulty or ease of performing a given
behavioral action~\cite{Lachen20}. Again, the TPB's behavioral control can be
interpreted in the insider threat context to align with the opportunities in the
SCP. The analytical meaning of the TBP (and TRA) is illustrated in
Fig.~\ref{fig: tbp}.

The social pressure in the TRA and the behavioral control in the TPB further
pinpoint toward two other theories, the so-called \textit{social bond theory}
(SBT) and the \textit{social learning theory}. The SBT posits that strong social
bonds may prevent an offender from committing a crime, despite the offender's
inclination to commit the crime. In reverse: bonding with criminals or other
misbehavers may increase the probability that an insider will commit an
offense. Such bonding is the message from the social learning theory; a person
is more likely to commit a crime if he or she associates with those who do so or
those who transmit delinquent ideas~\cite{Azaria14, Dupuis16}. Therefore, both
theories can be seen to fall into a domain of a more general sociological
\textit{social identity theory}; a person tends to behave in ways that comply
with the norms of formal and informal groups to which the person
belongs~\cite{Binns21}. The actual social bonds in the SBT are broken down into
four types: attachment (such as an affection and respect toward an
organization), commitment (such as an effort and energy to support the
organization's goals), involvement (such as a participation in organizational
activities), and beliefs (such as values and views about the
organization)~\cite{Safa18}. Strengthening such bonds should then reduce the
probability of insider incidents in an organization. The actual measures may
range from improving workplace culture to increasing the efficiency of human
resource management and seeking responsible leadership. Such measures and the
resulting bonds are closely related to the trust-related theorizing in the
literature~\cite{Gill17, MartinezMoyano06, Munshi12}. Despite technical and
organizational security measures, a management should ideally trust its
employees and the other way around; on one hand, a high-trust organizational
environment is likely to decrease the probability of insider incidents. On the
other hand, abuses of trust by people in positions of trust are a typical reason
behind insider incidents. This trust conundrum is not necessarily easy to
address because trust takes a long time to develop and restrictive security
controls may decrease trust among staff. This point is implicitly behind a
further theory, the so-called \textit{social exchange theory}~(SET) according to
which employees reciprocate their employer's treatment of
them~\cite{Moore18}. Thus: if an employee is mistreated, he or she is likely to
misbehave.

Finally, there is a general and rather straightforward \textit{ecological
  theory} in sociology, criminology, and associated social sciences: a person's
past behavior is likely to affect his or her future
behavior~\cite{Binns21}. Among other things, this theory justifies much of
recruitment practices; past educational and employment history together with
formative experiences are central to most hiring decisions. In criminology and
criminal law, the theory is implicitly also behind a long-lasting debate on
reductionism in criminal justice systems; there are no easy answers to a
question whether general crime prevention and intervention should be based on
past criminal behavior and associated profiling~\cite{Prins18}. With this point
in mind, the personal characteristics of insiders can be considered alongside
with the personal situations in which insiders find themselves.

\subsection{Personal Characteristics and Personal Situations}\label{sec: characteristics}

The aforementioned theories supplement presumptions about the personal
characteristics of an insider offender and his or her personal situation. Both
can be seen as more or less objective facts, not psychological assumptions about
a person, although both are likely to also influence the person's psychological
state and the other way around. In line with the ecological theory, formal
background checks done by law enforcement or intelligence agencies almost always
involve checking a person's criminal record~\cite{Binns21}. Clearly, then,
having an existing (cyber) crime record can be seen to increase the probability
that a person will commit an offense~\cite{Othman22}. Another example: having a
Ph.D.~in cyber security is also an objective fact about a person's
characteristics. Such a qualification may then correlate with the opportunity in
the SCP and maybe the motive too; the person in question may have (inside or
outside) knowledge on how to evade an organization's technical security controls
and hide his or her tracks.

Indeed, the literature reviewed tends to agree that technical knowledge, skills,
and competencies are relevant factors for insider threats~\cite{Ahmad14,
  Faresi11, Kisenasamy22, Marbut24, Munshi12, Nurse14, Sanders19}. In fact, many
documented insider incidents have involved employees in technical professions,
such as system administrators, database operators, or
programmers~\cite{Munshi12}. Technical competency tends to also reduce the
effectiveness of deterrence measures~\cite{Darcy09}. To these ends, some authors
have augmented the SCP to include also the capability of an insider to perform
an attack~\cite{Elmrabit20}, while others have considered skills in conjunction
with the opportunities~\cite{Farahmand13}. There are also other theoretical and
practical loose ends with respect to competencies and capabilities. For
instance, technical knowledge and skills may correlate with a person's
psychological traits, such as curiosity and aspiration for exploration, which,
in turn, may constitute a motive for a cyber crime~\cite{Prabhu22}. Another
point is that highly qualified people may be more familiar with an
organization's security policy. To use the Ph.D.~example again: a person with
such a qualification is likely also better aligned toward ethical guidelines and
professional conduct. Of course, such a presumption is dependent on a
context. For instance, when compared to a blue-collar worker, a scientist may be
more prone to steal intellectual property in order to start a
business~\cite{AlTabash18}. This example would connote with a motive to conduct
a misconduct as well as the personal characteristics.

Regarding even more fundamental personal characteristics, the literature is
surprisingly silent about basic demographic factors. Only in passing are such
factors rarely mentioned in the literature. For instance, it has been mentioned
that male employees in senior positions are more likely to commit
offenses~\cite{Othman22}, although also the contrary has been partially
observed; older employees are less likely to become malicious
insiders~\cite{Harms22, Prabhu22}. In~addition to gender and seniority,
ethno-cultural factors have been mentioned~\cite{Sokolowski16}. These factors
have also been mentioned with respect to susceptibility to phishing and social
engineering attacks~\cite{Dalal22}, which, as said, are also relevant for
insider threats. These fairly uncommon remarks notwithstanding, a rigorous and
systematic evaluation of demographic factors is absent in the literature
surveyed.

A further point is that the literature has often considered only technical
capabilities and competencies, omitting social and other ``softer''
skills. According to the SBT, such skills are likely to increase bonding with an
organization, which should reduce the probability of a misconduct. To this end,
job engagement and bonding with coworkers have been
considered~\cite{Moore18}. There are also studies that have examined symbolic
interactions in groups and their relation to the trustworthiness of the persons
involved~\cite{Ho14}. In addition, social isolation and remote work have been
considered as risk factors~\cite{Kisenasamy22}. These studies align, either
explicitly or implicitly, with the SBT's basic theoretical premises. In
addition, there are studies that are perhaps closer to the SET than the SBT. For
instance, some studies have considered organizational culture, ethics, and
organizations' support for employees to do their work as pull-off factors for
insider threats~\cite{Sokolowski16}. Here, particularly the organizational
support is a good example on the reciprocity assumption behind the
SET. Furthermore, there are studies that align, again either explicitly or
implicitly, with the TPB and its behavioral attitudes. A good example would be
job satisfaction and its relation to insider threats~\cite{Axelrad13,
  Sepehrzadeh23}. Also other attitudes, ideologies, subjective norms, and values
have been considered, such as patriotism, civil disobedience, disloyalty, and
dislike of authorities~\cite{Azaria14, Dalal22, Kisenasamy22,
  Marbut24}. However, the literature offers no cues on how the bonding,
reciprocity, attitudes, and related factors are related to a person's
characteristics, such as his or her people skills. Furthermore, personal
characteristics, like personality traits, are mostly static, whereas the factors
mentioned are rather dynamic. Therefore, also the motives are often not static
but instead develop or change over time~\cite{Munshi12, Sokolowski16}. For
instance, a person's job satisfaction may increase over time, which should imply
that the probability of he or she becoming a malicious insider decreases over
time. Likewise, bonding with coworkers takes time. A further example would
relate to the SET. If a person is continuously mistreated, the probability may
progressively increase that he or she will eventually commit a misconduct.

\begin{wrapfigure}{r}{0.45\textwidth}
\centering
\includegraphics[width=\linewidth, height=3cm]{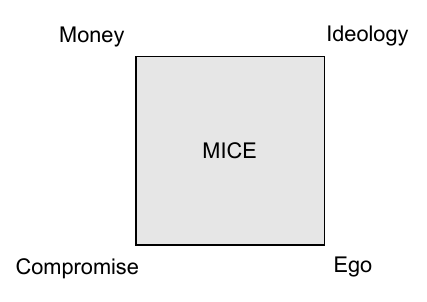}
\caption{The MICE Acronym}
\label{fig: mice}
\end{wrapfigure}

The dynamic nature of the factors mentioned correlate with a personal situation
of an insider; an employee continuously finds himself or herself in new
situations during his or her career and life in general. Here, the literature
has been particularly keen to examine different ``trigger factors'' that prompt
an insider to attack an organization from inside. A term precipitating events
has also been used to describe such triggers~\cite{Nurse14}. Such events or
trigger factors justify the term pressure in the fraud diamond; there is often
not only a motive but a specific pressure for a person to commit a
misconduct. Although not part of the literature reviewed, the so-called MICE
acronym is illuminating in this context (see Fig.~\ref{fig: mice}). It is a
mnemonic identifying some major factors that make a person prone to recruitment
of foreign intelligence services: money, ideology, compromise, and
ego~\cite{Petkus10}. As already said, ideology and politics may motive and
trigger some insiders, but according to the literature, money is a more typical
factor. The financial situation of a person is also an objective fact.

Indeed, the typical pressures considered in the insider threat literature are
financial problems and vices that a person have, a person's family situation,
termination of his or her employment contract, and negative organizational
changes~\cite{AlTabash18, Dupuis16, Faresi11, Marbut24, Othman22,
  Whitelaw24}. According to the literature, many insider incidents have occurred
soon after a person has left an organization either through a resignation or a
layoff; therefore, both contract termination dates and contract expiration dates
have been used as proxy variables for predicting insider
threats~\cite{AlTabash18, Elmrabit20}. Also the type of a contract (full-time or
part-time) has been considered~\cite{Faresi11}.  These variables likely
correlate with a financial pressure a person may have. In other words, a heavily
indebted employee with a part-time contract termination in sight may be a good
candidate for becoming a malicious insider. The negative organizational changes
may include a high staff turnover~\cite{Elmrabit20}, wage
reductions~\cite{AlTabash18, Dupuis16}, or an assignment of persons to new but
undesired work roles, among other things. In addition, the literature has
considered negative evaluations, corrective actions, and warnings as
predictors~\cite{Dupuis16, Faresi11, Greitzer22, Munshi12}. Although only seldom
contemplated, these factors and the associated pressures or triggers may also
change a person's psychological state. Before considering such states, further
more or less static elements of human beings should be elaborated.

\subsection{Personality Traits}\label{sec: traits}

Many of the papers sampled have considered different personality traits as
relevant factors for predicting insider threats or considering risks thereto. A
personality trait is a relatively stable habitual pattern of behavior, thought,
and emotion. Depending on a theory and operationalization, a personality trait
can be dichotomous, meaning that a person either has or has not a given trait,
or these can be seen to operate in an interval or continuous scale, such as when
the endpoints correspond with extraversion and introversion. Table~\ref{tab:
  traits} shows a summary of the traits considered in the literature
sampled. The summary does not mean that a given paper would necessarily operate
with a given trait empirically; many papers have also considered traits as
theoretical building blocks or illustrative examples on psychological factors in
the insider threat context. Nor does the summary include all traits considered
in the literature. As soon discussed, some were omitted due to theoretical and
other problems in the literature.

\begin{table}[th!b]
\centering
\caption{Examples of Personality Traits Considered in the Literature}
\label{tab: traits}
\begin{tabular}{ll}
\toprule
Trait & Papers \\
\hline
\multicolumn{2}{l}{\underline{The ``big five'' traits:}} \\
Conscientiousness & \cite{Alhajjar22, Axelrad13, Basu18, Brdiczka12, Brown13, Chi16, Duan24, Dupuis16, Eftimie21, Marbut24, Ren20, Schoenherr22, Sepehrzadeh23, Sticha16, Whitelaw24, Yang18, Yousef23} \\
Agreeableness & \cite{Alhajjar22, Axelrad13, Basu18, Brdiczka12, Brown13, Chi16, Duan24, Dupuis16, Eftimie21, Marbut24, Ren20, Sepehrzadeh23, Sticha16, Whitelaw24, Yang18, Yousef23} \\
Neuroticism  & \cite{Alhajjar22, Axelrad13, Basu18, Brdiczka12, Brown13, Chi16, Duan24, Dupuis16, Eftimie21, Marbut24, Ren20, Sepehrzadeh23, Sticha16, Whitelaw24, Yang18, Yousef23} \\
Openness & \cite{Alhajjar22, Basu18, Brdiczka12, Chi16, Duan24, Dupuis16, Eftimie21, Marbut24, Ren20, Sepehrzadeh23, Whitelaw24, Yang18, Yousef23} \\
Extraversion & \cite{Alhajjar22, Basu18, Brdiczka12, Chi16, Duan24, Dupuis16, Eftimie21, Marbut24, Ren20, Sepehrzadeh23, Whitelaw24, Yang18, Yousef23} \\
\multicolumn{2}{l}{\underline{The ``dark triad'' traits:}} \\
Narcissism & \cite{Basu18, Chi16, Harms22, Harrison18, Marbut24, Mathews17, Sanders19, Whitelaw24, Yang18} \\
Machiavellianism & \cite{Basu18, Chi16, Harms22, Harrison18, Marbut24, Sanders19, Whitelaw24, Yang18} \\
Psychopathy & \cite{Basu18, Chi16, Harms22, Harrison18, Marbut24, Sanders19, Whitelaw24, Yang18} \\
\multicolumn{2}{l}{\underline{Other dark traits:}} \\
Hostility & \cite{Dupuis16, Harms22, Marbut24} \\
Manipulativeness & \cite{Greitzer22, Harms22} \\
Deceitfulness & \cite{Harms22} \\
Anti-sociality & \cite{Marbut24} \\
Sadism & \cite{Marbut24} \\
\multicolumn{2}{l}{\underline{Other traits or sub-dimensions:}} \\
Impulsiveness & \cite{Harms22, Marbut24, Mathews17, Roy24} \\
Honesty and humility & \cite{Harms22, Marbut24, Schoenherr22} \\
Self-assurance & \cite{Dupuis16, Yang18} \\
Dutifulness & \cite{Harms22, Yang18} \\
Fearfulness & \cite{Dupuis16, Marbut24} \\
Empathy & \cite{Marbut24, Sarkar10} \\
Entitlement & \cite{Basu18, Marbut24} \\
Excitement-seeking \quad\quad & \cite{Axelrad13, Marbut24} \\
Self-centeredness & \cite{Azaria14, Greitzer22} \\
Excitableness & \cite{Harms22} \\
Skepticism & \cite{Harms22} \\
Callousness & \cite{Greitzer22} \\
Cautiousness & \cite{Harms22} \\
Reservedness & \cite{Harms22} \\
Leisureness & \cite{Harms22} \\
Boldness & \cite{Harms22} \\
Mischievousness & \cite{Harms22} \\
Colorfulness & \cite{Harms22} \\
Imaginativeness & \cite{Harms22} \\
Diligentness & \cite{Harms22} \\
Sympathy & \cite{Yang18} \\
Resilience & \cite{Marbut24} \\
Joviality & \cite{Dupuis16} \\
Attentiveness & \cite{Dupuis16} \\
Shyness & \cite{Dupuis16} \\
\bottomrule
\end{tabular}
\end{table}

As could be expected, much of the literature has operated with the so-called
``big five'' personality traits. These traits, which were initiated already in
the late 1950s but which gained prominence in the early
1990s~~\cite{Goldberg93}, are the \textit{de~facto} ones used in the
contemporary empirical literature, regardless of a discipline. Also the
so-called ``dark triad'', as pioneered in the early 2000s~\cite{Paulhouse02},
has been quite frequently used or discussed in the literature reviewed. This
triad is composed of three offensive but non-pathological personality types:
machiavellianism, narcissism, and psychopathy. Of these and other ``dark
traits'', it has been suspected that particularly non-pathological psychopathy
might be a good predictor for insider threats, although many of the darker
personality traits can be seen also as strong predictors of performance in cyber
security jobs~\cite{Harms22, Sanders19}. Non-pathological machiavellianism and
psychopathy are reportedly also decent predictors for determining engagement in
fraudulent behavior more generally~\cite{Harrison18}. The literature has also
considered traits that align directly with the MICE acronym discussed
earlier. In other words, a big ego and a markedly self-centered personality have
been perceived as risk factors for insider threats~\cite{Azaria14, Greitzer22,
  Prabhu22}. In addition, the literature has operated with numerous other
personality traits, some of which are problematic or even questionable, as soon
elaborated.

There are many problems in the literature empirically dealing with the
personality traits or otherwise operating with them. A few such problems can be
briefly noted. To begin with, many papers use particularly the big five
traits without any theoretical justifications on why these or other traits
should correlate with insider threats~\cite{Basu18, Chi16, Duan24, Dupuis16,
  Ren20, Yousef23}. There are three probable reasons for this atheoretical tenet
in the literature. The first reason is partly historical and partly related to
the availability of data: the big five traits have long been a part of the
popular insider threat datasets released by the CERT Coordination Center at the
Carnegie-Mellon university. The second reason is disciplinary: because most of
the papers are in the domain of computer science, rich psychological theorizing
is not expected, unlike perhaps machine learning and related applications. The
third reason is computational: particularly the big five traits are easily
available through scientific libraries or ready-made online frameworks, such as
the IBM's Watson artificial intelligence platform.

The computational aspects lead to a further problem: it is difficult to deduce
about the validity of the traits used in the literature because some papers have
used surveys \cite{Basu18, Dupuis16, Schoenherr22}, sometimes with validated
psychometric scales~\cite{Harms22, Harrison18}, and even professional
personality testing services~\cite{Eftimie21}, while others have relied on text
mining methods. In terms of the latter, particularly email datasets and
occasionally social networking data have been used together with sentiment
analysis~\cite{Brdiczka12, Brown13, Chi16, Yang18}. The methodological
uncertainties gain more weight in case personality traits are used in real-world
situations. Besides privacy issues, legal obstacles, and organizational ethics,
poorly conducted personality tests, likely including those done via machine
learning, perhaps without awareness of the employees involved, may put people
into unfavorable or unpleasant situations. Organizational trust may be involved
too, among other things; a person wrongly assigned as having psychopathic
traits, for instance, may no longer trust the given organization. Alternatively:
either current or prospective employees are not likely to answer candidly to
questions involving particularly the dark triad traits~\cite{Sanders19}. This
point reiterates the inherent methodological problems.

In terms of academic research and the literature reviewed, there are three
additional methodological and theoretical problems worth noting. The first
problem is that some studies have included only some of the big five traits. As
these are typically not statistically independent, multicollinearity may provide
a justification, but a theoretical rationale for omitting some traits is
typically lacking~\cite{Axelrad13, Brown13}. Some papers speak about a
``difficult personality factor'', as composed of neuroticism, conscientiousness,
and agreeableness, based on previous studies that have indicated that the traits
mentioned correlate with anti-social behavior~\cite{Sticha16}. While anti-social
behavior is in line with the SBT's social bonding assumptions, it remains
unclear why openness and extraversion would not be relevant for predicting
insider threats. Indeed, on one hand, some studies have included
excitement-seeking, which is a facet of extraversion, as a predictor for insider
threats~\cite{Axelrad13, Marbut24}. As noted in the previous section, curiosity
and exploration, which likely correlate with excitement-seeking, are often
considered as relevant when investigating motives for cyber crimes. On the other
hand, some papers have found that openness, extraversion, and agreeableness are
relevant predictors, but with a reverse interpretation; people who are creative,
social, and helpful to others are less likely to commit insider
offenses~\cite{Alhajjar22}. These conflicting interpretations underline the
literature's problems in theorizing.

The second problem is closely related: already the big five traits contain
numerous sub-dimensions of personality. Hence, it is unclear whether some
``traits'' considered in the literature really are personality traits. The
examples include trust, fear, guilt, anger, sadness, sympathy or empathy,
morality, altruism, and dutifulness~\cite{Brown13, Dupuis16, Marbut24,
  Yang18}. Of these, at least empathy and trust in others belong to the
agreeableness trait~\cite{Axelrad13}, and amorality either to the
machiavellianism or psychopathy trait~\cite{Harrison18}. The same goes with
factors such as concealment of things from others~\cite{Sticha16}, which might
be a sub-dimension of openness. Furthermore, some of the ``traits'' mentioned,
such as fear, guilt, sadness, and anger, are clearly emotions, not stable
personality traits of a person. The situation becomes even more complex once the
collusion between insiders and external threats is considered; already many of
the big five traits correlate with personal vulnerabilities to social
engineering attacks~\cite{Eftimie21, Greitzer14, Uebelacker14}. Thus, as has
been argued also in the literature reviewed~\cite{Harms22}, due to the inherent
complexity, including statistical problems, ``less might be more'' when
operating with personality traits.

The third problem follows: some studies have included ``traits'' that are
psychological disorders rather than conventional personality traits. The
examples include depression, borderline personality disorder, paranoia, and
disruptive mood dysregulation disorder~\cite{Ahmad14, Marbut24}. To put aside
the question whether these might be considered traits of a person's personality,
validity remains a big issue with such factors. As correctly noted in the
literature, these and related disorders or ``traits'' would require a clinical
diagnosis~\cite{Axelrad13}.  The literature has also discussed questionable
``disorders'' present in popular discourse, such as FOMO (fear of missing out)
and problematic Internet use~\cite{Mathews17}. Like previously with ``traits''
that are not necessarily traits, it remains unclear whether these ``disorders''
really are disorders. Furthermore, it also remains unclear how useful or
plausible such factors are in practical settings, including an organization's
human resource management, risk analysis procedures, and potential insider
threat predictions.

Thus, all in all, it is difficult to make systematic theoretical sense of the
literature operating with personality traits. It suffices to conclude that
personality traits do correlate with a likelihood of insider offenses, but the
theoretical reasons remain undecipherable. There is also the more fundamental
debate in psychology over the validity of the big five and other traits to
begin with. In addition to widespread measurement issues, personality traits
typically also vary across other factors, such as age, gender, and cultural
settings~\cite{Harms22}. The critical points raised align with other critical
viewpoints expressed in the literature surveyed. None of the personality traits
are inherently good or bad, and many of these also correlate with job
performance, security awareness, and so-called cyber hygiene
practices~\cite{Schoenherr21, Schoenherr22}. Therefore, instead of engaging in
intrusive psychological profiling with problematic methodologies and potentially
dangerous false positives, it might be a good idea for organizations to use
psychology for better motivating employees, including with respect to cyber
security practices~\cite{Munshi12}. This final point aligns well with the
general theoretical premises in the SBT and SET.

\subsection{Psychological States}\label{sec: states}

The literature reviewed has thus far addressed theoretical foundations, personal
characteristics, personal situations, and personality traits. Of these, personal
characteristics are relatively stable and rather objective facts about a person,
such his or her educational qualifications, expertise and employment history, or
a criminal record. Also personality traits are rather stable and long-term
habitual patterns; a person may be bold and extrovert, and such fundamental
personality traits unlikely change substantially and rapidly through the
person's life. In contrast, personal situations are dynamic and may change quite
rapidly. While financial problems may pile up over the years, it is also
possible that a person with a gambling habit loses a large amount of money in
one night. When compared to personal characteristics and personality traits, a
dismissal of a person from a job is also a more dynamic sequence; he or she may
receive prior warnings, but the firing event may still come as a surprise. It is
also possible that a whole company suddenly goes bankrupt and all its employee
are dismissed without prior warnings. These examples again illustrate the
pressures, triggers, and precipitating events discussed earlier. In what
follows, particularly these pressures, triggers, and events are discussed in
relation to a person's ``psychological state''.

Before continuing further, three additional brief remarks are in order. The
first is about the literature reviewed: there are some papers that have
considered ``psychological profiles'', ``psychological features'', or some
analogous constructs, but without proving any details on what such profiles or
features contain and on which theoretical and methodological premises they rely
on~\cite{Bishop09, Gamachchi15, LiLi22}. Clearly, it is impossible to engage
with such papers any further. The second point is about terminology: the concept
of a psychological state is not well-established in psychology. While there
seems to be no rigorous definitions, the literature reviewed has understood
psychological states to correspond with a person's psychological make-up and his
or her emotional state, both of which may change as a result of an
environment~\cite{Nurse14}. The make-up includes personality traits as well as
diagnosed disorders, such as, say, clinical depression. It should be
acknowledged that a given behavioral pattern may express different psychological
states at different times, and different persons may share the same
psychological state~\cite{Block72}. In other words, two depressed people may
have different personality traits and other personal characteristics, and the
same behavioral pattern that led to a person's diagnosis may not lead to a
further diagnosis. In any case, as psychological make-ups were already addressed
to the extent possible with the literature reviewed, emotions are what is left
to tackle. The third point follows: again, there are no rigorous definitions,
but an emotion can be generally understood as a physical and mental state
associated with thoughts, feelings, and behavioral responses. It is worth
picking an insight from the phenomenology of emotions; these are almost always
directed toward something~\cite{Szanto20}. While some emotions may be
self-directed, others are directed toward other people or objects. A~further
point is about the dynamics of emotions; these may develop and manifest
themselves over a long period of time, such as perhaps with a love of one's
special other, or these may manifest themselves as sudden outbursts, such as
might be the case with, say, outrage over something or some object. In the
insider threat context the direction toward which emotions are targeted may be a
given organization itself or the human beings working in the organization, such
as coworkers or executives. Emotions may be directed also toward clients,
customers, stakeholders, or other related parties of an organization.

With these preliminary points in mind, Table~\ref{tab: emotions} shows a summary
of typical emotions considered in the literature. As previously with the
personality traits, the summary is neither fully complete nor does it imply that
a given paper would operate with a given emotion empirically or otherwise
discuss it in detail. It is also worth remarking that rather similar conceptual
and theoretical issues are present than with personality traits. For instance,
revenge has been listed as an emotion~\cite{Prabhu22}, although it is a
behavioral pattern involving a commitment of a harmful action in response to
real or perceived grievances. In any case, revenge has been seen as a frequent
motive for insider attacks; in some papers it even surpasses a motive of
financial gains or a motive around opportunities~\cite{Marbut24}. Revenge is
usually also an emotionally laden hostile action taken by a person.

\begin{table}[th!b]
\centering
\caption{Examples of Emotions Considered in the Literature}
\label{tab: emotions}
\begin{tabular}{ll}
\toprule
Emotion & Papers \\
\hline
Stress & \cite{Azaria14, Carroll14, Dupuis16, Elmrabit20, Greitzer14, Harms22, Khan22, Mekonnen15, Moore18, Nurse14, Prabhu22, Renaud24, Sticha16, Whitelaw24} \\
Disgruntlement \quad\quad & \cite{Azaria14, Dupuis16, Greitzer22, Moore18, Nurse14, Prabhu22, Renaud24, Sarkar10, Sokolowski16, Sticha16, Whitelaw24} \\
Anger & \cite{Ahmad14, Azaria14, Dupuis16, Greitzer22, Nurse14, Prabhu22, Renaud24, Sticha16, Yang18} \\
Frustration & \cite{Nurse14, Prabhu22} \\
Anxiety & \cite{Greitzer14, Yang18} \\
Fear & \cite{Nurse14} \\
Boredom & \cite{Nurse14} \\
Jealousy & \cite{Whitelaw24} \\
Shame & \cite{Farshadkhah21} \\
Guilt & \cite{Farshadkhah21} \\
\bottomrule
\end{tabular}
\end{table}

Among the frequently discussed emotions is disgruntlement toward an
organizations or people working in it. It might be due to a denied promotion, a
lack of recognition, a mistreatment, or some other organizational
reason~\cite{Greitzer22, Renaud24}. Therefore, organizations should generally
seek to avoid unnecessary provocations in sensitive workplace matters, as these
may arouse disgruntlement and other strong negative
emotions~\cite{Safa19}. Then, the literature usually assumes that disgruntlement
in particular motives a revenge toward an organization, which, in turn,
eventually leads to an attack from inside the organization. Though, some papers
assume that disgruntlement first leads to counterproductive behavior before an
actual attack takes place~\cite{Sticha16}. Such counterproductive behavior might
include browsing job search portals or sending complaints to supervisors and
coworkers via electronic mail~\cite{Kan23}, or it may include generally
confrontational behavior~\cite{Azaria14}, among other things. Nevertheless,
disgruntlement remains the primary emotional trigger in this line of thought and
research.

Anger and stress are further commonly discussed emotions in the literature. To
put aside the question whether and how much anger and disgruntlement are
related, both are good examples for six reasons. First, stress, a feeling of
emotional strain and pressure, does not come out from the blue sky but develops
over time. Therefore, it correlates with other factors, such as work
performance~\cite{Carroll14, Greitzer14}, which, in turn, may correlate with job
evaluations and a person's attitudes toward an organization. A similar point
applies also to other emotions. For instance, a disgruntled employee may
under-perform in his or her work, which may cause poor job evaluations or even
lead to disciplinary action, which, in turn, may cause the employee to become
even more disgruntled~\cite{Nurse14}. Another example would relate to
exploration; as an insider engages in a transgression, his or her curiosity
increases, which may make it more difficult for him or her to stop his or her
misbehavior~\cite{Zaytsev17}. Although presumably difficult to empirically
observe, such vicious cycles are likely a risk factor for insider threats. In
general, however, the literature has not considered the intensification of
particular emotions beyond the examples mentioned.

The second point follows: nor has the literature considered the fact that
particular emotions may lead to further, different emotions. To use stress again
as an example: a heavily stressed employee may ask supervisors to reduce his or
her workload, but if his or her request is declined, the employee may become
angry toward the supervisors or the organization in general. If the situation
lasts for a long period of time, the stress and anger may then lead to
disgruntlement, which may lead to ideas about a revenge, which may motive an
insider attack. This kind of a vicious cycle resembles the reciprocity
assumption in the SET. Of course, it may also be that stress alone is a trigger
factor for an insider attack.

The third point is closely related: the emotions expressed by other people are
likely to affect a person's own emotions. While stress as an emotion is
supposedly mostly self-directed, an employee may become angry or even
disgruntled if he or she also observes that coworkers express cheerfulness and
joy at work, perhaps due to lighter workloads, real or perceived. To some
extent, this point has been considered in the literature; an employee's
disposition is affected by the dispositions of other
employees~\cite{Sokolowski16}. Therefore, it is no wonder that coworkers who
have witnessed an (unintentional) insider incident have shown feelings of guilt,
embarrassment, and frustration~\cite{Khan22}. Broadly speaking, these points are
underneath the bonding assumptions in the SBT. The social learning theory
provides a further point: if a stressed employee only bonds with other stressed
employees, they may all eventually become jointly angry or express other
negative emotions.

The fourth point is related to the previous discussion: not only may emotions
arouse from other people or an environment, including a given organization, but
a person's own situation is also a source of emotions. Here, as could be
expected, the literature has pointed out that financial pressures and other
problems in a person's life increase the stress levels of the
person~\cite{Dupuis16, Moore18}. A similar point applies to other emotions. For
instance, a person may become angry at himself or herself or at other people due
to problems in his or her own life. Closer to insider threats, a dismissal of an
employee is likely to arouse negative feelings.

Fifth, low stress tolerance and poor anger management have been seen as risk
factors for insider threats~\cite{Azaria14, Marbut24}. The ability to handle
stress, anger, and other emotions is partially a learned ability and partially
related to a person's personality. Thus, stress levels have been observed to
correlate with personality traits, such as agreeableness, conscientiousness, and
narcissism~\cite{Yang18}. This point further complicates the predictions done
with personality traits; some traits may predict insider threats, but it may
also be that some traits, possibly even the same traits, improve a person's
stress tolerance and anger management, which should reduce the probability of
insider attacks either directly or due to the emotions relation to
disgruntlement. It may even be that personality traits should be used as
confounding factors toward emotions, which may be the actual source for insider
threat risks. Furthermore, some personality traits, such as the dark ones, tend
to better or only manifest themselves in high-stress
situations~\cite{Harms22}. Some traits may also correlate with a tendency to
feel guilt or shame, which may prevent an insider from conducting a
transgression even in case he or she has already planned
one~\cite{Farshadkhah21}. Together, these points add further weight to the
critical reflection presented in the previous section.

The sixth and last point is about the technical approaches for insider threat
detection: intrusive monitoring and profiling are likely to cause stress among
employees~\cite{Mekonnen15}. Also other negative side-effects may be present,
including alienation of employees, reduction in morale, and decreasing
creativity~\cite{Moore18}. This final point serves to highlight that the
technical approaches to insider threats are not without their own problems; in
some cases, they may even be counterproductive.

Finally, it should be mentioned that in addition to surveys and other
conventional research methods, sentiment analysis has again been prominent in
the literature for profiling emotions. Emails~~\cite{Jiang18, Jiang19, Legg17,
  Mittal23, Soh19}, visited web pages~\cite{Jiang18, Jiang19, Legg17}, and
social media~\cite{Osterritter21} have provided the typical sources for
empirical data. These approaches repeat the previous point: it remains unclear
how such intrusive profiling fits into the legal landscape, workplace culture,
and organizational ethics, and how potential employees might perceive the
profiling and what consequences their perceptions might have. Furthermore, there
is also the important question of how well sentiments correspond with actual
emotions.

\subsection{Games, Deception, and Neurology}\label{sec: games}

The literature contains also three branches of research that do not connect well
with the other papers and the themes presented in them, although all branches
are still to some extent related to psychology. Already due the SLR protocol
used, these branches need a brief elaboration.

The first branch is about different games. Unlike what might be expected, these
are not typically game-theoretic games but rather concrete games involving
different tests and cognitive puzzles performed under stressful
conditions~\cite{Basu18}. There are also studies investigating games for
detecting lying~\cite{Chi16}, simulating betrayals~\cite{Ho17}, and detecting
deceptive behavior from textual cues~\cite{Ho16}. Also behavioral data from
mainstream online games has been investigated~\cite{Brdiczka12}. The second
branch is closely related: there are some studies that have investigated the use
of deception and honeypots for detecting insider threats~\cite{Alohaly22,
  Basu18, Takabi17}. Although the connection to psychology remains implicit in
both branches, the themes discussed and investigated still presumably correlate
with psychological make-ups, including personality traits and emotions.

The last branch is about biometrics and neurology. In terms of the former, eye
tracking has been a popular choice for insider threat
detection~\cite{Yerdon22}. It has also been used together with neurological
approaches. For instance, eye tracking has been used in conjunction with
functional magnetic resonance imaging (fMRI) in order to test the reception of
security messages under emotional constraints~\cite{Anderson16}. Alternatively,
eye tracking has also been used in conjunction with electroen-cephalogram (EEG)
signals for detecting insider threats~\cite{Hashem17}. As has been correctly
observed in the literature, such neurological data is highly sensitive personal
data, which may thus be a subject for insider threats in
itself~\cite{Ienca18}. Also other security and privacy risks have been
acknowledged in the literature~\cite{ChoiZage12, Kritika24}. Furthermore, it
remains unclear how useful, ethical, and legal these approaches are in
practice. If psychological profiling is seen as intrusive, clearly brain
scanning is even more intrusive. Thus, it remains debatable whether current or
prospective employees would consent to fMRI or EEG scanning, and whether labor
and privacy laws would even allow such scanning at a workplace.

\section{Conclusion}\label{sec: conclusion}

What do we know about the psychology of insider threats? At first glance, the
answer might be: not much. On a second thought, the literature reviewed posits a
general picture in which personal characteristics, personal situations, and
other more or less objective facts correlate with personality traits,
psychological states, and different behavioral patterns. However, (a) the
literature suffers from a general lack of robust theorizing on how these and
related concepts are related to each other. As was elaborated in
Section~\ref{sec: theory}, there are plenty of theoretical references, and there
is nothing wrong in borrowing theories from other fields, but the connection of
these references to psychology remains more or less implicit. In particular,
causal relations are generally ambiguous and undertheorized.

Another research gap follows: (b) there is a general lack of validation and
replication studies as well as pre-registered studies. Although some attempts
have been made~\cite{Gheyas16}, this lack together with the absence of
comparable theories implies that meta-analysis is generally impossible. In other
words, (c)~it remains impossible to say which theories outperform other
theories. Even more importantly, (d)~it is also impossible to say whether the
technical approaches outperform the non-technical approaches, including those
based on psychology. As was discussed in Sections~\ref{sec: traits}, \ref{sec:
  states}, and~\ref{sec: games}, psychological and related profiling is
generally problematic in terms of privacy, ethics, workplace culture, and even
law. Even though psychology remains an important topic in academic insider
threat research, it would therefore be important to know whether the technical
approaches suffice alone for insider threat detection; that is, whether
psychological profiling is even needed in practical settings. In terms of
academic research, furthermore, with some rare exceptions~\cite{Alhajjar22},
(e)~the literature reviewed has mostly operated with static snapshots of
data. As was pointed out in Sections~\ref{sec: characteristics}~and~\ref{sec:
  states}, neither motives nor emotions are static; therefore, more longitudinal
research is generally needed in order to better understand insider threats and
their psychology. Finally, (f)~there are some methodological and related
problems in the literature. Among other things, proxy variables have been quite
loosely and liberally used for probing different psychological aspects of
persons. This criticism affects also machine learning approaches. For instance,
it remains debatable how well sentiment analysis can proxy actual emotions of
persons.

As has been pointed out also in other reviews~\cite{Gheyas16}, also other
conventional issues may be present, including so-called publication bias. Among
other things, (g)~negative results, including nonworking or implausible
methodological approaches, are also missing from the literature. Finally and
importantly, (h)~the SLR protocol used indicated no relevant papers published in
psychology~journals.

\bibliographystyle{splncs03}

\begin{thebibliography}{10}
\providecommand{\url}[1]{\texttt{#1}}
\providecommand{\urlprefix}{URL }

\bibitem{Ahmad14}
Ahmad, M.B., {Saeed-ur-Rehman}, Akram, A., Asif, M.: {T}owards a {R}ealistic
  {R}isk {A}ssessment {M}ethodology for {I}nsider {T}hreats of {I}nformation
  {M}isuse. In: Proceedings of the 12th International Conference on Frontiers
  of Information Technology. pp. 176--181. IEEE, Islamabad (2014)

\bibitem{AlTabash18}
{Al tabash}, K., Happa, J.: {I}nsider-{T}hreat {D}etection {U}sing {G}aussian
  {M}ixture {M}odels and {S}ensitivity {P}rofiles. Computers \& Security  77,
  838--859 (2018)

\bibitem{Alhajjar22}
Alhajjar, E., Bradley, T.: {S}urvival {A}nalysis for {I}nsider {T}hreat.
  Computational and Mathematical Organization Theory  28,  335--351 (2022)

\bibitem{Alohaly22}
Alohaly, M., Balogun, O., Takabi, D.: {I}ntegrating {C}yber {D}eception {I}nto
  {A}ttribute-{B}ased {A}ccess {C}ontrol {(ABAC)} for {I}nsider {T}hreat
  {D}etection. IEEE Access  10,  108965--108978 (2022)

\bibitem{Anderson16}
Anderson, B.B., Vance, A., Kirwan, C.B., Eargle, D., Jenkins, J.L.: {H}ow
  {U}sers {P}erceive and {R}espond to {S}ecurity {M}essages: a {N}euro{IS}
  {R}esearch {A}genda and {E}mpirical {S}tudy. European Journal of Information
  Systems  25,  364--390 (2016)

\bibitem{Axelrad13}
Axelrad, E.T., Sticha, P.J., Brdiczka, O., Shen, J.: {A} {B}ayesian {N}etwork
  {M}odel for {P}redicting {I}nsider {T}hreats. In: Proceedings of the IEEE
  Security and Privacy Workshops. pp. 82--89. IEEE, San Francisco (2023)

\bibitem{Azaria14}
Azaria, A., Richardson, A., Kraus, S., Subrahmanian, V.S.: {B}ehavioral
  {A}nalysis of {I}nsider {T}hreat: {A} {S}urvey and {B}ootstrapped
  {P}rediction in {I}mbalanced {D}ata. IEEE Transactions on Computational
  Social Systems  1(2),  135--155 (2014)

\bibitem{Basu18}
Basu, S., Chua, Y.H.V., Lee, M.W., Lim, W.G., Maszczyk, T., Guo, Z., Dauwels,
  J.: {T}owards a {D}ata-{D}riven {B}ehavioral {A}pproach to {P}rediction of
  {I}nsider-{T}hreat. In: Proceedings of the IEEE International Conference on
  Big Data (Big Data). pp. 4994--5001. IEEE, Seattle (2018)

\bibitem{Binns21}
Binns, C.A., Kempf, R.J.: {B}ackground {C}hecks: {T}he {T}heories {B}ehind the
  {P}rocess. Security Journal  34,  776--801 (2021)

\bibitem{Bishop09}
Bishop, M., Gates, C., Frincke, D., Greitzer, F.L.: {AZALIA}: {A}n {A} to {Z}
  {A}ssessment of the {L}ikelihood of {I}nsider {A}ttack. In: Proceedings of
  the IEEE Conference on Technologies for Homeland Security. pp. 385--392.
  IEEE, Waltham (2009)

\bibitem{Block72}
Block, N.J., Fodor, J.A.: {W}hat {P}sychological {S}tates {A}re {N}ot. The
  Philosophical Review  81(2),  159--181 (1972)

\bibitem{Brdiczka12}
Brdiczka, O., Liu, J., Price, B., Shen, J., Patil, A., Chow, R., Bart, E.,
  Ducheneaut, N.: {P}roactive {I}nsider {T}hreat {D}etection {T}hrough {G}raph
  {L}earning and {P}sychological {C}ontext. In: Proceedings of the IEEE
  Symposium on Security and Privacy Workshops. pp. 142--149. IEEE, San
  Francisco (2012)

\bibitem{Brown13}
Brown, C.R., Watkins, A., Greitzer, F.L.: {P}redicting {I}nsider {T}hreat
  {R}isks {T}hrough {L}inguistic {A}nalysis of {E}lectronic {C}ommunication.
  In: Proceedings of the 46th Hawaii International Conference on System
  Sciences. pp. 1849--1858. IEEE, Wailea (2013)

\bibitem{Carroll14}
Carroll, T.E., Greitzer, F.L., Roberts, A.D.: {S}ecurity {I}nformatics
  {R}esearch {C}hallenges for {M}itigating {C}yber {F}riendly {F}ire. Security
  Informatics  13,  1--14 (2014)

\bibitem{Chan08}
Chan, D.K.: {I}ntroduction: {M}oral {P}sychology {T}oday. In: Chan, D.K. (ed.)
  {M}oral {P}sychology {T}oday: {E}ssays on {V}alues, {R}ational {C}hoice, and
  the {W}ill. Springer, Cham (2008)

\bibitem{Chi16}
Chi, H., Scarllet, C., Prodanoff, Z.G., Hubbard, D.: {D}etermining
  {P}redisposition to {I}nsider {T}hreat {A}ctivities by {U}sing {T}ext
  {A}nalysis. In: Proceedings of the Future Technologies Conference (FTC). pp.
  985--990. IEEE, San Francisco (2016)

\bibitem{ChoiZage12}
Choi, S., Zage, D.: {A}ddressing {I}nsider {T}hreat {U}sing ``{W}here {Y}ou
  {A}re'' as {F}ourth {F}actor {A}uthentication. In: Proceedings IEEE
  International Carnahan Conference on Security Technology (ICCST). pp.
  147--153. IEEE, Newton (2012)

\bibitem{CISA23}
{CISA}: {Z}ero {T}rust {M}aturity {M}odel (2023), {C}ybersecurity \&
  {I}nfrastructure {S}ecurity {A}gency {(CISA)} of the {U}nited {S}tates.
  {A}vailable online in July 2024:
  \url{https://www.cisa.gov/sites/default/files/2023-04/zero_trust_maturity_model_v2_508.pdf}

\bibitem{Dalal22}
Dalal, R.S., Howard, D.J., Brummel, B.J.: {O}rganizational {S}cience and
  {C}ybersecurity: {A}bundant {O}pportunities for {R}esearch at the
  {I}nterface. Journal of Business and Psychology  37,  1--29 (2022)

\bibitem{Darcy09}
{D'Arcy}, J., Hovav, A.: {D}oes {O}ne {S}ize {F}it {A}ll? {E}xamining the
  {D}ifferential {E}ffects of {IS} {S}ecurity {C}ountermeasures. Journal of
  Business Ethics  89,  59--71 (2009)

\bibitem{Dong24}
Dong, B., Chernov, S., Akpina, K.O.: {L}egal {A}spects of {C}orporate {S}ystems
  for {P}reventing {C}ybercrime {A}mong {P}ersonnel. Crime, Law and Social
  Change  81,  75--96 (2024)

\bibitem{Duan24}
Duan, S., Yuan, J., Wang, B.: {C}ontextual {F}eature {R}epresentation for
  {I}mage-{B}ased {I}nsider {T}hreat {C}lassification. Computers \& Security
  140,  103779 (2024)

\bibitem{Dupuis16}
Dupuis, M., Khadeer, S.: {C}uriosity {K}illed the {O}rganization: {A}
  {P}sychological {C}omparison {B}etween {M}alicious and {N}on-{M}alicious
  {I}nsiders and the {I}nsider {T}hreat. In: Proceedings of the 5th Annual
  Conference on Research in Information Technology. pp. 35--40. ACM, Boston
  (2016)

\bibitem{Eftimie21}
Eftimie, S., Cotenescu, V., R\u{a}cuciu, C., Gl\u{a}van, D.: {A} {C}ase {S}tudy
  in {A}nticipating {I}nsider {V}ulnerabilities {U}sing {P}sychological
  {P}rofiling. In: Proceedings of the IEEE International Black Sea Conference
  on Communications and Networking (BlackSeaCom). pp. 1--4. IEEE, Bucharest
  (2021)

\bibitem{Elmrabit20}
Elmrabit, N., Yang, S., Yang, L., Zhou, H.: {I}nsider {T}hreat {R}isk
  {P}rediction {B}ased on {B}ayesian {N}etwork. Computers \& Security  96,
  101908 (2020)

\bibitem{Farahmand13}
Farahmand, F., Spafford, E.H.: {U}nderstanding {I}nsiders: {A}n {A}nalysis of
  {R}isk-{T}aking {B}ehavior. Information Systems Frontiers  15,  5--15 (2013)

\bibitem{Faresi11}
Faresi, A.A., Wijesekera, D.: {P}reemptive {M}echanism to {P}revent {H}ealth
  {D}ata {P}rivacy {L}eakage. In: Proceedings of the International Conference
  on Management of Emergent Digital EcoSystems. pp. 17--24. ACM, San Francisco
  (2011)

\bibitem{Farshadkhah21}
Farshadkhah, S., {Van Slyke}, C., Fuller, B.: {O}nlooker {E}ffect and
  {A}ffective {R}esponses in {I}nformation {S}ecurity {V}iolation {M}itigation.
  Computers \& Security  100,  102082 (2021)

\bibitem{Fishbein77}
Fishbein, M., Ajzen, I.: {B}elief, {A}ttitude, {I}ntention, and {B}ehavior:
  {A}n {I}ntroduction to {T}heory and {R}esearch. Philosophy and Rhetoric
  10(2),  130--132 (1977)

\bibitem{Gamachchi15}
Gamachchi, A., Bozta\c{s}, S.: {W}eb {A}ccess {P}atterns {R}eveal {I}nsiders
  {B}ehavior. In: Proceedings of the International Workshop on Signal Design
  and its Applications in Communications (IWSDA). pp. 70--74. IEEE, Bengaluru
  (2015)

\bibitem{Gheyas16}
Gheyas, I.A., Abdallah, A.E.: {D}etection and {P}rediction of {I}nsider
  {T}hreats to {C}yber {S}ecurity: {A} {S}ystematic {L}iterature {R}eview and
  {M}eta-{A}nalysis. Big Data Analytics  1(6),  1--29 (2016)

\bibitem{Gill17}
Gill, M., Crane, S.: {T}he {R}ole and {I}mportance of {T}rust: {A} {S}tudy of
  the {C}onditions that {G}enerate and {U}ndermine {S}ensitive {I}nformation
  {S}haring. Security Journal  30,  734--748 (2017)

\bibitem{Goldberg93}
Goldberg, L.R.: {T}he {S}tructure of {P}henotypic {P}ersonality {T}raits.
  American Psychologist  48(1),  26--34 (1993)

\bibitem{Greitzer22}
Greitzer, F.L., Purl, J.: {T}he {D}ynamic {N}ature of {I}nsider {T}hreat
  {I}ndicators. SN Computer Science  3,  1--15 (2022)

\bibitem{Greitzer14}
Greitzer, F.L., Strozer, J.R., Cohen, S., Moore, A.P., Mundie, D., Cowley, J.:
  {A}nalysis of {U}nintentional {I}nsider {T}hreats {D}eriving from {S}ocial
  {E}ngineering {E}xploits. In: Proceedings of the IEEE Security and Privacy
  Workshops. pp. 236--250. IEEE (2014)

\bibitem{Harms22}
Harms, P., Marbut, A., Johnston, A.C., Lester, P., Fezzey, T.: {E}xposing the
  {D}arkness {W}ithin: {A} {R}eview of {D}ark {P}ersonality {T}raits, {M}odels,
  and {M}easures and {T}heir {R}elationship to {I}nsider {T}hreats. Journal of
  Information Security and Applications  71,  103378 (2022)

\bibitem{Harrison18}
Harrison, A., Summers, J., Mennecke, B.: {T}he {E}ffects of the {D}ark {T}riad
  on {U}nethical {B}ehavior. Journal of Business Ethics  153,  53--77 (2018)

\bibitem{Hashem17}
Hashem, Y., Takabi, H., Dantu, R., Nielsen, R.: {A} {M}ulti-{M}odal
  {N}euro-{P}hysiological {S}tudy of {M}alicious {I}nsider {T}hreats. In:
  Proceedings of the 2017 International Workshop on Managing Insider Security
  Threats. pp. 33--44. ACM, Dallas (2017)

\bibitem{Hiebl23}
Hiebl, M.R.W.: {S}ample {S}election in {S}ystematic {L}iterature {R}eviews of
  {M}anagement {R}esearch. Organizational Research Methods  26(2),  229--261
  (2023)

\bibitem{Ho14}
Ho, S.M., Benbasat, I.: {D}yadic {A}ttribution {M}odel: {A} {M}echanism to
  {A}ssess {T}rustworthiness in {V}irtual {O}rganizations. Journal of the
  Association for Information Science and Technology  65(8),  1555--1576 (2014)

\bibitem{Ho16}
Ho, S.M., Hancock, J.T., Booth, C., Burmester, M., Liu, X., Timmarajus, S.S.:
  {D}emystifying {I}nsider {T}hreat: {L}anguage-{A}ction {C}ues in {G}roup
  {D}ynamics. In: Proceedings of the 49th Hawaii International Conference on
  System Sciences (HICSS). pp. 2729--2738. IEEE, Koloa (2016)

\bibitem{Ho17}
Ho, S.M., Warkentin, M.: {L}eader's {D}ilemma {G}ame: {A}n {E}xperimental
  {D}esign for {C}yber {I}nsider {T}hreat {R}esearch. Information Systems
  Frontiers  19,  377--396 (2017)

\bibitem{Ienca18}
Ienca, M., Haselager, P., Emanuel, E.J.: {B}rain {L}eaks and {C}onsumer
  {N}eurotechnology. Nature Biotechnology  36(9),  805--810 (2018)

\bibitem{Jiang18}
Jiang, J., Chen, J., Choo, K.R., Liu, K., Liu, C., Yu, M., Mohapatra, P.:
  {P}rediction and {D}etection of {M}alicious {I}nsiders' {M}otivation {B}ased
  on {S}entiment {P}rofile on {W}ebpages and {E}mails. In: Proceedings of the
  IEEE Military Communications Conference (MILCOM). pp. 1--6. IEEE, Los Angeles
  (2018)

\bibitem{Jiang19}
Jiang, J., Chen, J., Gu, T., Choo, K.R., Liu, C., Yu, M., Huang, W., Moha, P.:
  {W}arder: {O}nline {I}nsider {T}hreat {D}etection {S}ystem {U}sing
  {M}ulti-{F}eature {M}odeling and {G}raph-{B}ased {C}orrelation. In:
  Proceedings of the IEEE Military Communications Conference (MILCOM). pp.
  1--6. IEEE, Norfolk (2019)

\bibitem{Kan23}
Kan, X., Fan, Y., Zheng, J., Kudreyko, A., Chi, C., Song, W., Tregubova, A.:
  {U}ser-{L}evel {M}alicious {B}ehavior {A}nalysis {M}odel {B}ased on the
  {NMF-GMM} {A}lgorithm and {E}nsemble {S}trategy. Nonlinear Dynamics  111,
  21391--21408 (2023)

\bibitem{Khan22}
Khan, N., Houghton, R.J., Sharples, S.: {U}nderstanding {F}actors {T}hat
  {I}nfluence {U}nintentional {I}nsider {T}hreat: {A} {F}ramework to
  {C}ounteract {U}nintentional {R}isks. Cognition, Technology \& Work  24,
  393--421 (2022)

\bibitem{Kisenasamy22}
Kisenasamy, K., Perumal, S., Raman, V., Singh, B.S.M.: {I}nfluencing {F}actors
  {I}dentification in {S}mart {S}ociety for {I}nsider {T}hreat in {L}aw
  {E}nforcement {A}gency {U}sing a {M}ixed {M}ethod {A}pproach. International
  Journal of System Assurance Engineering and Management  13(Suppl 1),
  236--251 (2022)

\bibitem{Kitchenham13}
Kitchenham, B., Brereton, P.: {A} {S}ystematic {R}eview of {S}ystematic
  {R}eview {P}rocess {R}esearch in {S}oftware {E}ngineering. Information and
  Software Technology  55,  2049--2075 (2013)

\bibitem{Koien19}
K\o{}ien, G.M.: {W}hy {C}ryptosystems {F}ail {R}evisited. Wireless Personal
  Communication  106,  85--117 (2019)

\bibitem{Kritika24}
Kritika, M.: {A} {C}omprehensive {S}tudy on {N}avigating {N}euroethics in
  {C}yberspace. AI and Ethics (Published oline in May),  1--8 (2024)

\bibitem{Lachen20}
Lahcen, R.A.M., Caulkins, B., Mohapatra, R., Kumar, M.: {R}eview and {I}nsight
  on the {B}ehavioral {A}spects of {C}ybersecurity. Cybersecurity  3,  1--18
  (2020)

\bibitem{LeeLallie23}
Lee, D., Lallie, H.S., Michaelides, N.: {T}he {I}mpact of an {E}mployee's
  {P}sychological {C}ontract {B}reach on {C}ompliance with {I}nformation
  {S}ecurity {P}olicies: {I}ntrinsic and {E}xtrinsic {M}otivation. Cognition,
  Technology \& Work  25,  273--289 (2023)

\bibitem{Legg17}
Legg, P.A., Buckley, O., Goldsmith, M., Creese, S.: {A}utomated {I}nsider
  {T}hreat {D}etection {S}ystem {U}sing {U}ser and {R}ole-{B}ased {P}rofile
  {A}ssessment. IEEE Systems Journal  11(2),  503--512 (2017)

\bibitem{LiLi22}
Li, C., Li, F., Yu, M., Guo, Y., Wen, Y., Li, Z.: {I}nsider {T}hreat
  {D}etection {U}sing {G}enerative {A}dversarial {G}raph {A}ttention
  {N}etworks. In: Proceedings of the IEEE Global Communications Conference
  (GLOBECOM). pp. 2680--2685. IEEE, Rio de Janeiro (2022)

\bibitem{Marbut24}
Marbut, A.R., Harms, P.D.: {F}iends and {F}ools: {A} {N}arrative {R}eview and
  {N}eo‑{S}ocioanalytic {P}erspective on {P}ersonality and {I}nsider
  {T}hreats. Journal of Business and Psychology  39,  679--696 (2024)

\bibitem{MartinezMoyano06}
{Martinez-Moyano}, I.J., Rich, E.H., Conrad, S.H., Andersen, D.F.: {M}odeling
  the {E}mergence of {I}nsider {T}hreat {V}ulnerabilities. In: Proceedings of
  the Winter Simulation Conference. pp. 562--568. IEEE, Monterey (2006)

\bibitem{Mathews17}
Mathews, R.: {I}nterrogating ``{P}rivacy'' in a {W}orld {B}rimming with {H}igh
  {P}olitical {E}ntanglements, {S}urveillance, {I}nterdependence \&
  {I}nterconnections. Health and Technology  7,  265--324 (2017)

\bibitem{Mekonnen15}
Mekonnen, S., Padayachee, K., Meshesha, M.: {A} {P}rivacy {P}reserving
  {C}ontext-{A}ware {I}nsider {T}hreat {P}rediction and {P}revention {M}odel
  {P}redicated on the {C}omponents of the {F}raud {D}iamond. In: Proceedings of
  the Annual Global Online Conference on Information and Computer Technology
  (GOCICT). pp. 60--65. IEEE, Louisville (2015)

\bibitem{Mittal23}
Mittal, A., Garg, U.: {P}rediction and {D}etection of {I}nsider {T}hreat
  {D}etection {U}sing {E}mails: {A} {C}omparision. In: Proceedings of the
  Second International Conference on Electrical, Electronics, Information and
  Communication Technologies (ICEEICT). pp. 1--6. Trichirappalli (2023)

\bibitem{Moore18}
Moore, A.P., Cassidy, T.M., Theis, M.C., Bauer, D., Rousseau, D.M., Moore,
  S.B.: {B}alancing {O}rganizational {I}ncentives to {C}ounter {I}nsider
  {T}hreat. In: Proceedings of the IEEE Security and Privacy Workshops (SPW).
  pp. 237--246. IEEE (2018)

\bibitem{Munshi12}
Munshi, A., Dell, P., Armstrong, H.: {I}nsider {T}hreat {B}ehavior {F}actors:
  {A} {C}omparison of {T}heory {W}ith {R}eported {I}ncidents. In: Proceedings
  of the 45th Hawaii International Conference on System Sciences. pp.
  2402--2411. IEEE, Maui (2021)

\bibitem{Nightingale09}
Nightingale, A.: {A} {G}uide to {S}ystematic {L}iterature {R}eviews. Surgery
  27(9),  381--384 (2009)

\bibitem{Nurse14}
Nurse, J.R.C., Buckley, O., Legg, P.A., Goldsmith, M., Creese, S., Wright,
  G.R.T., Whitty, M.: {U}nderstanding {I}nsider {T}hreat: {A} {F}ramework for
  {C}haracterising {A}ttacks. In: Proceedings of the IEEE Security and Privacy
  Workshops. pp. 214--228. IEEE, San Jose (2014)

\bibitem{Osterritter21}
Osterritter, L., Carley, K.M.: {C}onversations {A}round {O}rganizational {R}isk
  and {I}nsider {T}hreat. In: Proceedings of the 2021 IEEE/ACM International
  Conference on Advances in Social Networks Analysis and Mining. pp. 613--621.
  ACM (2021)

\bibitem{Othman22}
Othman, R., Ameer, R.: {I}n {E}mployees {W}e {T}rust: {E}mployee {F}raud in
  {S}mall {B}usinesses. Journal of Management Control  33,  189--213 (2022)

\bibitem{Paulhouse02}
Paulhouse, D.L., Williams, K.M.: {T}he {D}ark {T}riad of {P}ersonality:
  {N}arcissism, {M}achiavellianism, and {P}sychopathy. Journal of Research in
  Personality  36(6),  556--563 (2002)

\bibitem{Petersen08}
Petersen, K., Feldt, R., Mujtaba, S., Mattson, M.: {S}ystematic {M}apping
  {S}tudies in {S}oftware {E}ngineering. In: Proceedings of the 12th
  International Conference on Evaluation and Assessment in Software Engineering
  (EASE). pp. 68--77. BCS Learning \& Development Ltd., Italy (2008)

\bibitem{Petkus10}
Petkus, D.D.: {E}thics of {H}uman {I}ntelligence {O}perations: {O}f {MICE} and
  {M}en. International Journal of Intelligence Ethics  1(1),  97--121 (2010)

\bibitem{Prabhu22}
Prabhu, S., Thompson, N.: {A} {P}rimer on {I}nsider {T}hreats in
  {C}ybersecurity. Information Security Journal: A Global Perspective  31(5),
  602--611 (2022)

\bibitem{Prins18}
Prins, S.J., Reich, A.: {C}an {W}e {A}void {R}eductionism in {R}isk
  {R}eduction? Theoretical Criminology  22(2),  258--278 (2018)

\bibitem{Ren20}
Ren, X., Wang, L.: {A} {H}ybrid {I}ntelligent {S}ystem for {I}nsider {T}hreat
  {D}etection {U}sing {I}terative {A}ttention. In: Proceedings of 2020 6th
  International Conference on Computing and Data Engineering. pp. 189--194. ACM
  (2020)

\bibitem{Renaud24}
Renaud, K., Warkentin, M., Pogrebna, G., {van der Schyff}, K.: {VISTA}: {A}n
  {I}nclusive {I}nsider {T}hreat {T}axonomy, {W}ith {M}itigation {S}trategies.
  Information \& Management  61,  103877 (2024)

\bibitem{Roy24}
Roy, K.C., Chen, G.: {G}raph{CH}: {A} {D}eep {F}ramework for {A}ssessing
  {C}yber-{H}uman {A}spects in {I}nsider {T}hreat {D}etection. IEEE
  Transactions on Dependable and Secure Computing (Preprint),  1--15 (2024)

\bibitem{Ruohonen24PRR}
Ruohonen, J., Hjerppe, K., Kortesuo, K.: {C}risis {C}ommunication in the {F}ace
  of {D}ata {B}reaches (2024), {A}rchived manuscript. Available online in June:
  \url{https://arxiv.org/abs/2406.01744}

\bibitem{Ruohonen24IWCC}
Ruohonen, J., Hjerppe, K., von Zastrow, M.: {A}n {E}xploratory {C}ase {S}tudy
  on {D}ata {B}reach {J}ournalism. In: Proceedings of the 19th International
  Conference on Availability, Reliability and Security (ARES 2024). pp. 1--9.
  ACM, Vienna (2024)

\bibitem{Safa19}
Safa, N.S., Maple, C., Furnell, S., Azad, M.A., Perera, C., Dabbagh, M.,
  Sookhak, M.: {D}eterrence and {P}revention-{B}ased {M}odel to {M}itigate
  {I}nformation {S}ecurity {I}nsider {T}hreats in {O}rganisations. Future
  Generation Computer Systems  97,  587--597 (2019)

\bibitem{Safa18}
Safa, N.S., Maple, C., Watson, T., {Von Solms}, R.: {M}otivation and
  {O}pportunity {B}ased {M}odel to {R}educe {I}nformation {S}ecurity {I}nsider
  {T}hreats in {O}rganisations. Journal of Information Security and
  Applications  40,  247--257 (2018)

\bibitem{Sanders19}
Sanders, G.L., Upadhyaya, S., Wang, X.: {I}nside the {I}nsider. IEEE
  Engineering Management Review  47(2),  84--91 (2019)

\bibitem{Santos12}
Santos, E., Nguyen, H., Yu, F., Kim, K.J., Li, D., Wilkinson, J.T., Olson, A.,
  Russell, J., , Clark, B.: {I}ntelligence {A}nalyses and the {I}nsider
  {T}hreat. IEEE Transactions on Systems, Man, and Cybernetic -- Part A:
  Systems and Humans  42(2),  331--347 (2012)

\bibitem{Sarkar10}
Sarkar, K.R.: {A}ssessing {I}nsider {T}hreats to {I}nformation {S}ecurity
  {U}sing {T}echnical, {B}ehavioural and {O}rganisational {M}easures.
  Information Security Technical Report  15,  112--133 (2010)

\bibitem{Schoenherr22}
Schoenherr, J.R.: {I}nsider {T}hreats and {I}ndividual {D}ifferences:
  {I}ntention and {U}nintentional {M}otivations. IEEE Transactions on
  Technology and Society  3(3),  175--184 (2022)

\bibitem{Schoenherr21}
Schoenherr, J.R., Thomson, R.: {T}he {C}ybersecurity {(CSEC)} {Q}uestionnaire:
  {I}ndividual {D}ifferences in {U}nintentional {I}nsider {T}hreat
  {B}ehaviours. In: Proceedings of the International Conference on Cyber
  Situational Awareness, Data Analytics and Assessment (CyberSA). pp. 1--8.
  IEEE, Dublin (2021)

\bibitem{Sepehrzadeh23}
Sepehrzadeh, H.: {A} {M}ethod for {I}nsider {T}hreat {A}ssessment by {M}odeling
  the {I}nternal {E}mployee {I}nteractions. International Journal of
  Information Security  22,  1385--1393 (2023)

\bibitem{Soh19}
Soh, C., Yu, S., Narayanan, A., Duraisamy, S., Chen, L.: {E}mployee
  {P}roﬁling via {A}spect-{B}ased {S}entiment and {N}etwork for {I}sider
  {T}hreats {D}etection. Expert Systems With Applications  135,  351--361
  (2019)

\bibitem{Sokolowski16}
Sokolowski, J.A., Banks, C.M., Dover, T.J.: {A}n {A}gent-{B}ased {A}pproach to
  {M}odeling {I}nsider {T}hreat. Computational and Mathemtical Organization
  Theory  22,  273--287 (2016)

\bibitem{Sokolowski15}
Sokolowski, J.A., Banks, C.M.: {A}n {A}gent-{B}ased {A}pproach to {M}odeling
  {I}nsider {T}hreat. In: Proceedings of the Symposium on Agent-Directed
  Simulation. pp. 36--41. ACM, San Diego (2015)

\bibitem{Sticha16}
Sticha, P.J., Axelrad, E.T.: {U}sing {D}ynamic {M}odels to {S}upport
  {I}nferences of {I}nsider {T}hreat {R}isk. Computational and Mathemtical
  Organization Theory  22,  350--381 (2016)

\bibitem{Szanto20}
Szanto, T., Landweer, H.: {I}ntroduction: {T}he {P}henomenology of
  {E}motions---{A}bove and {B}eyond '{W}hat {I}t {I}s {L}ike to {F}eel'. In:
  Szanto, T., Landweer, H. (eds.) {T}he {R}outledge {H}andbook of
  {P}henomenology of {E}motion, pp. 1--37. Routledge, Oxford (2020)

\bibitem{Takabi17}
Takabi, H., Jafarian, J.H.: {I}nsider {T}hreat {M}itigation {U}sing {M}oving
  {T}arget {D}efense and {D}eception. In: Proceedings of the 2017 International
  Workshop on Managing Insider Security Threats. pp. 93--96. ACM (2017)

\bibitem{Uebelacker14}
Uebelacker, S., Quiel, S.: {T}he {S}ocial {E}ngineering {P}ersonality
  {F}ramework. In: Proceedings of the Workshop on Socio-Technical Aspects in
  Security and Trust. pp. 24--30. IEEE, Vienna (2014)

\bibitem{Whitelaw24}
Whitelaw, F., Riley, J., Elmrabit, N.: {A} {R}eview of the {I}nsider {T}hreat,
  a {P}ractitioner {P}erspective {W}ithin the {U.K.} {F}inancial {S}ervices.
  IEEE Access  12,  34752--34768 (2024)

\bibitem{Willison06}
Willison, R., Backhouse, J.: {O}pportunities for {C}omputer {C}rime:
  {C}onsidering {S}ystems {R}isk from a {C}riminological {P}erspective.
  European Journal of Information Systems  15,  403--414 (2006)

\bibitem{Yang18}
Yang, G., Cai, L., Yu, A., Ma, J., Meng, D., Wu, Y.: {P}otential {M}alicious
  {I}nsiders {D}etection {B}ased on a {C}omprehensive {S}ecurity
  {P}sychological {M}odel. In: Proceedings of the IEEE Fourth International
  Conference on Big Data Computing Service and Applications (BigDataService).
  pp. 9--16. Bamberg (2018)

\bibitem{Yerdon22}
Yerdon, V.A., Lin, J., Wohleber, R.W., Matthews, G., {Reinerman-Jones}, L.,
  Hancock, P.A.: {E}ye-{T}racking {A}ctive {I}ndicators of {I}nsider {T}hreats:
  {D}etecting {I}llicit {A}ctivity {D}uring {N}ormal {W}orkflow. IEEE
  Transactions on Engineering Management  69(6),  3838--3847 (2022)

\bibitem{Yousef23}
Yousef, R., Jazzar, M., Eleyan, A., Bejaoui, T.: {A} {M}achine {L}earning
  {F}ramework \& {D}evelopment for {I}nsider {C}yber-{C}rime {T}hreats
  {D}etection. In: Proceedings of the International Conference on Smart
  Applications, Communications and Networking (SmartNets). pp. 1--6. IEEE,
  Istanbul (2023)

\bibitem{Zaytsev17}
Zaytsev, A., Malyuk, A., Miloslavskaya, N.: {C}ritical {A}nalysis in the
  {R}esearch {A}rea of {I}nsider {T}hreats. In: Proceedings of the IEEE 5th
  International Conference on Future Internet of Things and Cloud (FiCloud).
  pp. 288--296. IEEE, Prague (2017)

\end{thebibliography}

\end{document}